# Technological Competence is a Precondition for Effective Implementation of Virtual Reality Head Mounted Displays in Human Neuroscience: A Technological Review and Meta-analysis.

**Panagiotis Kourtesis[1,2,3,4], Simona Collina[3,4], Leonidas A.A. Doumas[2], and Sarah E. MacPherson[1,2]**

[1] Human Cognitive Neuroscience, Department of Psychology, University of Edinburgh, Edinburgh, United Kingdom

[2] Department of Psychology, University of Edinburgh, Edinburgh, United Kingdom

[3] Lab of Experimental Psychology, Suor Orsola Benincasa University of Naples, Naples, Italy

[4] Interdepartmental Centre for Planning and Research "Scienza Nuova", Suor Orsola Benincasa University of Naples, Naples, Italy

*** Correspondence:**
Panagiotis Kourtesis
pkourtes@exseed.ed.ac.uk

**Keywords: Virtual Reality1, VRISE2, HMD3, Cybersickness4, Neuroscience5, Neuropsychology6, Psychology7, VR8.**

**Abstract**

Immersive virtual reality (VR) emerges as a promising research and clinical tool. However, several studies suggest that VR induced adverse symptoms and effects (VRISE) may undermine the health and safety standards, and the reliability of the scientific results. In the current literature review, the technical reasons for the adverse symptomatology are investigated to provide suggestions and technological knowledge for the implementation of VR head-mounted display (HMD) systems in cognitive neuroscience. The technological systematic literature indicated features pertinent to display, sound, motion tracking, navigation, ergonomic interactions, user experience, and computer hardware that should be considered by the researchers. Subsequently, a meta-analysis of 44 neuroscientific or neuropsychological studies involving VR HMD systems was performed. The meta-analysis of the VR studies demonstrated that new generation HMDs induced significantly less VRISE and marginally fewer dropouts. Importantly, the commercial versions of the new generation HMDs with ergonomic interactions had zero incidents of adverse symptomatology and dropouts. HMDs equivalent to or greater than the commercial versions of contemporary HMDs accompanied with ergonomic interactions are suitable for implementation in cognitive neuroscience. In conclusion, researchers' technological competency, along with meticulous methods and reports pertinent to software, hardware, and VRISE, are paramount to ensure the health and safety standards and the reliability of neuroscientific results.

## 1 Introduction

In recent years, virtual reality (VR) technology has attracted attention, demonstrating its utility and potency in the field of neuroscience and neuropsychology (Rizzo *et al.*, 2004; Bohil *et al.*, 2011; Parsons, 2015). Traditional approaches in human neuroscience involve the utilization of static and simple stimuli which arguably lack ecological validity (Parsons, 2015). VR offers the usage of dynamic stimuli and interactions with a high degree of control within an ecologically valid environment which enables the collection of advanced cognitive and behavioral data (Rizzo *et al.*, 2004; Bohil *et al.*, 2011; Parsons, 2015). VR can be combined with non-invasive imaging techniques (Bohil *et al.*, 2011; Parsons, 2015) and has been effective in the assessment of cognitive and affective functions and clinical conditions (e.g., social stress disorders) which require ecological validity (Rizzo *et al.*, 2004; Parsons, 2015) for their assessment, rehabilitation and treatment (e.g., post-traumatic stress disorder) (Rizzo *et al.*, 2004; Bohil *et al.*, 2011).

However, researchers and clinicians have reported caveats with the implementation of immersive VR interventions and assessments, particularly when head mounted display (HMD) systems are utilized (Sharples *et al.*, 2008; Davis *et al.*, 2015; de Franca & Soares, 2017; Palmisano *et al.*, 2017). A predominant concern is the presence of adverse physiological symptoms (i.e., cyber/simulation-sickness which includes nausea, disorientation, instability, dizziness, and fatigue). These undesirable effects are categorized as VR Induced Symptoms and Effects (VRISE) (Sharples *et al.*, 2008; Davis *et al.*, 2015; de Franca & Soares, 2017; Palmisano *et al.*, 2017), and are evaluated by using questionnaires such as the Simulator Sickness Questionnaire (Kennedy *et al.*, 1993) and the Virtual Reality Sickness Questionnaire (Kim *et al.*, 2018).

VRISE may risk the health and safety of participants or patients (Kane & Parsons, 2017; Parsons *et al.*, 2018), which raises ethical considerations for the adoption of VR HMDs as research and clinical tools. Additionally, the presence of VRISE has modulated substantial decline in reaction times and overall cognitive performance (Plant & Turner, 2009; Nalivaiko *et al.*, 2015; Plant, 2016; Nesbitt *et al.*, 2017; Mittelstaedt *et al.*, 2018), as well as increasing body temperature and heart rates (Nalivaiko *et al.*, 2015). Also, the presence of VRISE robustly increases cerebral blood flow and oxyhemoglobin concentration (Gavgani et al, 2018), the power of brain signals (Arafat *et al.*, 2018), and the connectivity between stimulus response brain regions and nausea-processing brain regions (Toschi *et al.*, 2017). Thus, VRISE could be considered confounding variables, which significantly undermine the reliability of neuropsychological, physiological, and neuroimaging data.

VRISE are predominantly mediated by an oculomotor discrepancy between what is being perceived through the oculomotor (optic nerve) sensor and what is being sensed via the rest of the afferent nerves in the human body (Sharples *et al.*, 2008; Davis *et al.*, 2015; de Franca & Soares, 2017; Palmisano *et al.*, 2017). Nevertheless, technologically speaking, VRISE are derivatives of hardware and software inadequacies, i.e., the type of display screen, resolution and refresh rate of the image, the size of the field of view as well as non-ergonomic movements within an interaction in the virtual environment (VE; de Franca & Soares, 2017; Palmisano *et al.*, 2017). Notably, VR HMDs have substantially evolved during the last two decades. Important differences may be seen between the HMDs released before 2013 (old generation) and those released from 2013 onwards (new generation). While the last old generation HMD was released in 2001 (i.e., nVisor SX111), the year 2013 is used to distinguish between old and new generation HMDs, since it is the year that the first new generation HMD prototype (i.e., Oculus Development Kit 1) was released. This systematic review attempts to clarify the technological etiologies of VRISE and provide pertinent suggestions for the implementation of VR HMDs in cognitive neuroscience and neuropsychology. In addition, a meta-analysis of the neuroscience studies that have implemented VR HMDs will be conducted to elucidate the frequency of VRISE and dropout rates as per the VR HMD generation.

## 2 Technological systematic review

In Table 1, a glossary of the key terms and concepts is provided to assist with comprehension of the utilized terminology. We followed the Preferred Reporting Items for Systematic Reviews and Meta-Analyses (PRISMA) guidelines using a decremental stepwise method to perform the literature review (see Figure 1). The selected papers and book chapters included an explicit explanation and discussion of VRISE and users' experiences pertinent with the specified technological features of the VR hardware and software. Digital databases specialized in technologies were used: 1) IEEE Xplore Digital Library; 2) ACM Digital Library; 3) ScienceDirect; 4) MIT CogNet; and 5) Scopus. Two categories of keywords were used, where each category had three or more keywords and each paper had to include at least one keyword from each category in the main body of the text. The categories were: 1) "virtual reality" OR "immersive virtual reality" OR "head-mounted display"; AND 2) "VRISE" OR "motion sickness" OR "cyber sickness" OR "simulation sickness". Finally, the extracted information from the identified papers was clustered together under common features (i.e., display, sound, motion tracking, navigation, ergonomic interactions, user experience, and computer hardware).

Table 1. Glossary of key terms and concepts

| | Terms & Concepts | Explanation/Definition |
|---|---|---|
| Headsets | Head Mounted Display (HMD) | A display device which is worn on the head and provides an immersive virtual reality for the wearer. |
| | Development Kit (DK) HMD | A prototype device, which is utilized by the VR Software developers to develop VR software before the commercial version of an HMD. The DKs are not provided for general use. |
| | Commercial Version (CV) HMD | The final version of an HMD, which is dispersed to the market for general use. |
| Display | Liquid Crystal Display (LCD) | A type of display/screen that uses the light-modulating properties of liquid crystals. Liquid crystals emit light indirectly, instead of using a reflector to produce images. |
| | Organic Light Emitting Diode (OLED) | A type of display/screen that uses an organic compound film that emits light in response to an electric current. OLEDs are used as displays in devices such as television screens, computer monitors, and smartphones. |
| | Field of View (FOV) | The area captured by the display device. The size of the FOV and the size of the display device directly affect the quality of the image. |
| | Refresh Rate & Frame Rate | The refresh rate is the number of times that the hardware updates its display per second. It involves the repeated display of identical frames. The frame rate indicates the frequency that software can add new data to a display. |
| | Resolution | The number of distinct pixels in each dimension displayed in a frame. |
| Interactions | Motion tracking | The process of tracking the movement of objects or people. It is facilitated by motion sensors which detect the position of motion trackers embedded in devices (e.g., HMDs and 6 DoF controllers). |
| | Controllers/Wands with 6 degrees of freedom (DoF) | Controllers which have 6DoF of movement in 3-D space on three directional axes (i.e., Forward-Back, Left-Right, Up-Down) and three rotational axes (i.e., Roll, Pitch, Yaw). |
| | Direct Hand Interaction | A motion tracking device (i.e., a motion sensor) which directly tracks hand movements |
| | Teleportation | A navigation system, which allows the user to be transferred to a new location in the virtual environment without physically moving in the real environment. |
| | Ergonomic Interactions | These resemble real-life interactions, which optimize user experience and overall VR system performance (see also 3.2.3 Definition of Ergonomic Interactions). Ergonomic interactions are facilitated by and restricted to the capabilities of the VR hardware and software. |
| | Virtual Environment (VE) | A three-dimensional artificial environment which is displayed on a display device and allows the users to interact with it. |

Figure 1. Decremental stepwise method for the technological literature review.

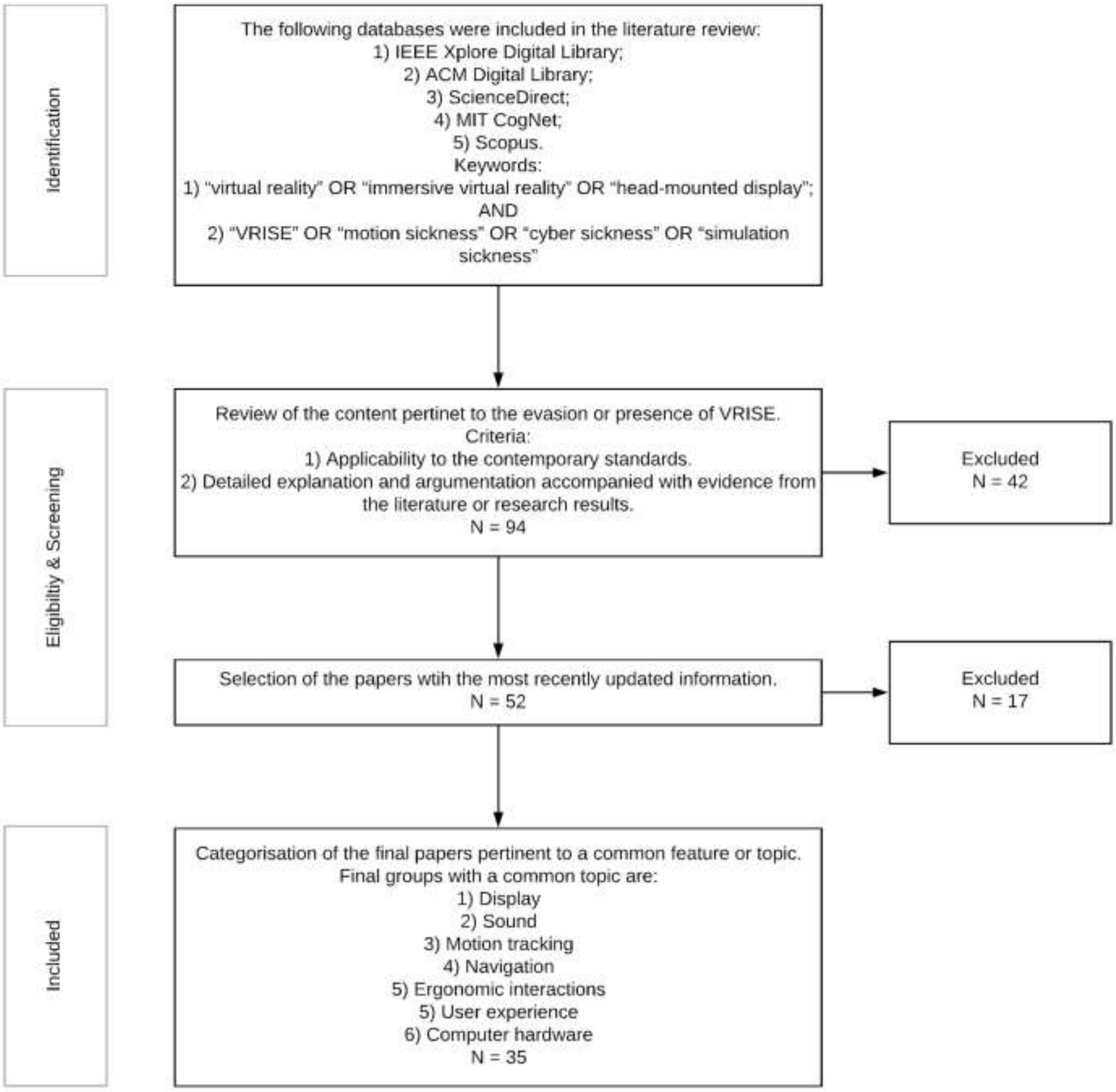

## 2.1 Technological Etiologies of VRISE

### 2.1.1 Display

VR HMDs use the following three types of screens: Cathode Ray Tubes (CRT); Liquid Crystal Display (LCD); and organic light emitting diode (OLED). LCD screens replaced CRT ones due to VRISE (Costello, 1997). LCD, in comparison to CRT, alleviated the probability of visual complications and physical burdens (e.g., fatigue) (Costello, 1997). However, the suitability of LCD was challenged by the emergence of OLED screens. While old generation VR HMDs mainly utilize LCD screens (Costello, 1997), the commercial versions of new generation VR HMDs predominantly use OLED screens (Kim *et al.*, 2017). The OLED screens have been found to be better than LCD screens for general implementation in VR, because of their faster response times, lighter weight, and better color quality (Kim *et al.*, 2017). OLED screens decrease the likelihood of VRISE and offer an improved VR display (Kim *et al.*, 2017).

Three more factors related to display type are crucial for the avoidance of VRISE: the width of the Field of View (FOV) (Rakkolainen *et al.*, 2016; Kim *et al.*, 2017); the resolution of the image per eye (Hecht, 2016; Rakkolainen *et al.*, 2016; Kim *et al.*, 2017; Brennesholtz, 2018); and the latency of the images (frames per second) (Hecht, 2016; Rakkolainen *et al.*, 2016; Kim *et al.*, 2017; Brennesholtz, 2018). A wider FOV significantly decreases the chance of VRISE and increases the level of immersion (Rakkolainen *et al.*, 2016; Kim *et al.*, 2017). The canonical guidelines suggest a lowest threshold of $110^{o}$ FOV (diagonal) (Hecht, 2016; Rakkolainen *et al.*, 2016; Kim *et al.*, 2017; Brennesholtz, 2018). In addition, an increased refresh rate and resolution alleviates the danger of discomfort or VRISE (Hecht, 2016; Rakkolainen *et al.*, 2016; Kim *et al.*, 2017; Brennesholtz, 2018). The refresh rate should be $\geq$ 75Hz (i.e., $\geq$ 75 frames per second) (Goradia *et al.*, 2014; Hecht, 2016; Brennesholtz, 2018), while the resolution is required to be higher than 960 x 1080 sub-pixels per eye (Goradia *et al.*, 2014).

### 2.1.2 Sound

A second important consideration for a user's experience in VR is the sound quality. The integration of spatialized sounds (e.g., ambient and feedback sounds) in the VE may increase the level of immersion, pleasantness of the experience, and successful navigation (Vorländer & Shinn-Cunningham, 2014), while they significantly decrease the likelihood of VRISE (Viirre *et al.*, 2014). However, the volume and localization of sounds need to be optimized in terms of audio spatialization to ensure a user's experience is pleasant without adverse VRISE (Vorländer & Shinn-Cunningham, 2014; Viirre *et al.*, 2014).

### 2.1.3 Motion tracking

Motion tracking in VR is a precondition for naturalistic movement within an immersive VE (Slater & Wilbur, 1997; Stanney & Hale, 2014). Motion tracking allows the precise tracking of the user's physical body within the VE (i.e., it allows the computer to provide accurate environmental feedback, which modulates and consolidates the awareness of the position and movement of the user's body). This phenomenon is called proprioception or kinesthesia (Slater & Wilbur, 1997) and is linked with vestibular and oculomotor mediated VRISE (Slater & Wilbur, 1997; Plouzeau *et al.*, 2015; Caputo *et al.*, 2017). Hence, motion tracking should be adequately rapid and accurate to facilitate ergonomic interactions in the VE (Caputo *et al.*, 2017).

### 2.1.4 Navigation

A highly important factor in the quality of VR software and to avoid VRISE is the movement of the user in the VE (Porcino *et al.*, 2017). New generation HMDs deliver an adequate play area for interactions to facilitate ecologically valid scenarios (Porcino *et al.*, 2017; Borrego *et al.*, 2018). However, there are restrictions in the size of the play area, which does not permit navigation solely by physical walking (Porcino *et al.*, 2017; Borrego *et al.*, 2018). Teleportation allows movement beyond the play area size and elicits a high-level of immersion and pleasant user experience, whilst alleviating VRISE (Bozgeyikli *et al.*, 2016; Frommel *et al.*, 2017; Porcino *et al.*, 2017). In contrast, movement dependent on a touchpad, keyboard, or joystick results to high occurrences of VRISE (Bozgeyikli *et al.*, 2016; Frommel *et al.*, 2017; Porcino *et al.*, 2017). Therefore, teleportation in conjunction with physical movement (i.e., free movement of the upper limbs and walking in a small-restricted area) is the most suitable method for movement in VR (Bozgeyikli *et al.*, 2016; Frommel *et al.*, 2017; Porcino *et al.*, 2017). Yet, there are additional factors such as external hardware (i.e., controllers and wands), which are needed to facilitate optimal ergonomic interactions in VR.

### 2.1.5 Ergonomic Interactions

Ergonomic and naturalistic interactions are essential to minimize the risk of VRISE, while non-ergonomic and non-naturalistic interactions increase the occurrence of them (Slater & Wilbur, 1997; Stanney & Hale, 2014; Plouzeau *et al.*, 2015; Caputo *et al.*, 2017; Porcino *et al.*, 2017). Importantly, controllers, joysticks, and keyboards do not support ergonomic and naturalistic interactions in VR (Plouzeau *et al.*, 2015; Bozgeyikli *et al.*, 2016; Caputo *et al.*, 2017; Frommel *et al.*, 2017; Porcino *et al.*, 2017; Sportillo *et al.*, 2017; Figueiredo *et al.*, 2018). Instead, wands with 6 degrees of freedom (DoF) of movement (e.g., Oculus Rift and HTC Vive wands), and realistic interfaces with direct hand interactions (e.g., Microsoft's Kinect) facilitate naturalistic and ergonomic interactions (Sportillo *et al.*, 2017; Figueiredo *et al.*, 2018). Both hardware systems facilitate easy familiarization with their controls and their utilization (Sportillo *et al.*, 2017; Figueiredo *et al.*, 2018). However, direct hand interactions are easier than 6DoF controllers-wands in terms of familiarization with their controls and efficiency (Sportillo *et al.*, 2017; Figueiredo *et al.*, 2018). Direct hand interactions were also found to offer more pleasant user experiences (Sportillo *et al.*, 2017; Figueiredo *et al.*, 2018), although, they are substantially less accurate than 6DoF controller-wands (Sportillo *et al.*, 2017; Figueiredo *et al.*, 2018).

### 2.1.6 User Experience

Notably, ergonomic interactions might be available to the user; however, the user is required to learn the necessary interactions and how the VE functions to facilitate a pleasant user experience (Gromala *et al.*, 2016; Jerald *et al.*, 2017; Brade *et al.*, 2018). The inclusion of comprehensible tutorials where the user may spend an adequate amount of time acquiring the necessary skills (i.e., navigation, use and grab of items, two-handed interactions) and knowledge of the VE (i.e., how it reacts to your controls) is crucial (Gromala *et al.*, 2016; Jerald *et al.*, 2017; Brade *et al.*, 2018). Additionally, in-game instructions and prompts should be offered to the user through interactions in the VE (e.g., directional arrows, non-player characters, signs, labels, ambient sounds, audio, and videos) (Gromala *et al.*, 2016; Jerald *et al.*, 2017; Brade *et al.*, 2018).

### 2.1.7 Computer Hardware

The computer hardware (i.e., the processor, graphics card, sound card) should at least meet the minimum requirements of the VR software and HMD (Anthes *et al.*, 2016). The performance of VR HMDs is analogous to the computing power and the quality of the hardware (Anthes *et al.*, 2016;

Stanney & Hale, 2014; Borrego *et al.*, 2018). The processor, graphics card, sound card, and operating system (e.g., Windows) need to be considered and reported because they modulate the performance of the software (Plant & Turner, 2009; Plant, 2016; Kane & Parsons, 2017; Parsons *et al.*, 2018). Research software developers and researchers are required to be technologically competent in order to opt for the appropriate hardware and software to achieve their research and/or clinical aims (Plant, 2016; Kane & Parsons, 2017; Parsons *et al.*, 2018).

## 2.2 Conclusions

Based on the outcomes of the above technological review, VR HMDs should have a good quality display-screen (i.e., OLED or upgraded LCD), an adequate FOV (i.e., diagonal FOV $\geq 110^{o}$), adequate resolution per eye (i.e., resolution > 960 x 1080 sub-pixels per eye), and an adequate image refresh rate (i.e., refresh rate ≥75Hz) to safeguard the health and safety of the participants and the reliability of the neuroscientific results (see Table 2). Also, the VR HMD should have external hardware which offers an adequate VR area, fast and accurate motion tracking, spatialized audio, and ergonomic interactions. The computer's processor, graphics card, and sound card should meet the minimum requirements of the VR software and HMD too. New generation VR HMDs appear to have all the necessary hardware characteristics (i.e., graphics, level of immersion, and sound) to be used in ecological valid research and clinical paradigms (Borrego *et al.*, 2018; see Table 2 for a comparison between old and new generation HMDs). New generation VR HMDs have the required hardware to support and produce high-quality spatialized sounds in VEs (Borrego *et al.*, 2018). Additionally, new generation VR HMDs have integrated rapid and precise motion tracking which facilitates naturalistic and ergonomic interactions within the VE (Borrego *et al.*, 2018).

Both the Oculus development kit (DK) 1 and DK2 do not meet the minimum hardware features highlighted by the technological review, despite being new generation VR HMDs (see Table 2). The DK1 has substantially lower resolution per eye and image refresh rates, while the DK2 has marginally acceptable refresh rates, yet a slightly lower resolution per eye. These DKs are not available for general use but are used by professional developers to produce beta (early) versions of their games or apps (Goradia *et al.*, 2014; Suznjevic *et al.*, 2017). Moreover, they were removed from the market after the release of the Oculus Rift CV. VR HMDs should have hardware characteristics equal to or better than the commercial versions (CV) of the Oculus Rift and HTC Vive in order to ensure the health & safety of the participants, as well as the reliability of the neuroscientific results (i.e., physiological, neuropsychological, and neuroimaging data). The researchers and clinicians should have the technological competence to choose an HMD which is equal to or greater than the CVs of the Oculus Rift and HTC Vive (e.g., Valve Index, HTC Vive Pro, Oculus Quest, Pimax VR, and StarVR).

However, the VR software's features are equally important. The VR software should include an ergonomic interaction and navigation system, as well as tutorials, in-game instructions, and prompts. A suitable navigation system should combine teleportation and physical movement, while ergonomic interactions should include those that simulate real-life interactions by using a direct hands system or 6DoF controllers. Also, the tutorials, in-game instructions, and prompts should be informative and easy to follow, especially for experimental or clinical purposes where users should be equally able to interact with the VE (Plant & Turner, 2009; Plant, 2016; Kane & Parsons, 2017; Parsons *et al.*, 2018). The criteria for effective VR software are displayed in Table 3. These criteria should be met before implementing VR software for research and/or clinical purposes. Otherwise, researchers or clinicians may compromise the reliability of their study's results (Plant & Turner, 2009; Plant, 2016;

Kane & Parsons, 2017; Parsons *et al.*, 2018), and/or jeopardize the health and safety of their participants/patients (Kane & Parsons, 2017; Parsons *et al.*, 2018).

The above features enable researchers or clinicians to administer a sophisticated and pleasant VR experience, which substantially alleviates or eradicates adverse VRISE. Therefore, the technological competency of neuroscientists and neuropsychologists is a precondition for the efficient adoption and implementation of innovative technologies like VR HMDs in cognitive neuroscience or neuropsychology.

Table 2. Minimum hardware criteria: old and new generation VR HMDs

| Product | Generation | Resolution (per eye) | Display Screen | Refresh Rate | FOV (Diagonal) | Motion Trackers and Sensors (Type and Quantity) |
|---|---|---|---|---|---|---|
| VFX 3D | Old | 480 x 240 | LCD | 45 Hz | 45° | - |
| VUZIX Wrap 1200 | Old | 852x480 | LCD | 60 Hz | 35° | Unknown type (1), 3 magnetometers, 3 accelerometers, and 3 gyroscopes |
| eMagin Z800 3DVisor | Old | 800x600 | OLED | 60 Hz | 40° | - |
| nVisor SX111 | Old | 1280x1024 | LCD | 60 Hz | 110° | - |
| Oculus Rift Development Kit 1 | New | 640 x 800 | LCD | 60 Hz | 110° | - |
| Oculus Rift Development Kit 2 | New | 960 x 1080 | OLED | 75 Hz | 110° | - |
| Minimum Hardware Criteria for the Avoidance of VRISE | NA | > 960 x 1080 | OLED or LCD | ≥ 75Hz | ≥ 110° | Tracking should be adequately rapid and accurate to facilitate ergonomic interactions. |
| Oculus Rift Commercial Version | New | 1080 x 1200 | OLED | 90 Hz | 110° | Accelerometer, gyroscope, magnetometer, 360⁰ constellation tracking camera. |
| HTC VIVE Commercial Version | New | 1080 x 1200 | OLED | 90 Hz | 110° | Sensors (>70) including MEMS, magnetometer, gyroscope, accelerometer, and laser position sensors, Lighthouse laser tracking system (2 base stations emitting pulsed InfraRed lasers), front-facing camera. |

*MEMS = Microelectromechanical systems*

Table 3. Criteria for suitable VR software in cognitive neuroscience and neuropsychology

| Domains | User Experience | Game Mechanics | In-Game Assistance | VRISE |
|---|---|---|---|---|
| CRITERIA | An Adequate Level of Immersion | A Suitable Navigation System (e.g., Teleportation) | Digestible Tutorials | Absence or Insignificant Presence of Nausea |
| | Pleasant VR Experience | Availability of Physical Movement | Helpful Tutorials | Absence or Insignificant Presence of Disorientation |
| | High Quality Graphics | Naturalistic Picking/Placing of Items | Adequate Duration of Tutorials | Absence or Insignificant Presence of Dizziness |
| | High Quality Sounds | Naturalistic Use of Items | Helpful In-game Instructions | Absence or Insignificant Presence of Fatigue |
| | Suitable Hardware (HMD and Computer) | Naturalistic 2-Handed Interaction | Helpful In-game Prompts | Absence or Insignificant Presence of Instability |

# 3 Meta-analysis of VR studies in cognitive neuroscience

## 3.1 Literature Research and Inclusion Criteria

We followed the PRISMA guidelines to conduct the literature research using a decremental approach, where the selection commenced with a relatively vast accumulation of abstracts and concluded with a diminished list of full papers that comprise standardised and detailed VR research paradigms. The procedure is described in Figure 2. The following databases were used for the literature research: 1) PsycInfo; 2) PsycArticles; 3) PubMed; and 4) Medline. Two categories of keywords were used, with three keywords in each category. The minimum threshold for each study was the inclusion of at least one keyword from each category in the main body of text. The keywords for each category were: 1) "virtual reality"; OR "Immersive"; OR "Head Mounted Display"; AND 2) "Psychology"; OR "Neuropsychology" OR "Neuroscience". Additional filters and criteria were: 1) chronological specification (2004 and later); and 2) a comprehensive description of the VR research methods in conjunction with the research aims and results. Finally, the selected studies were allocated into two groups according to the generation of the implemented VR HMD. Two tables display the studies that utilise old generation (Table 4) and new generation (Table 5) HMDs.

Figure 2. Decremental stepwise method for the literature review of VR studies.

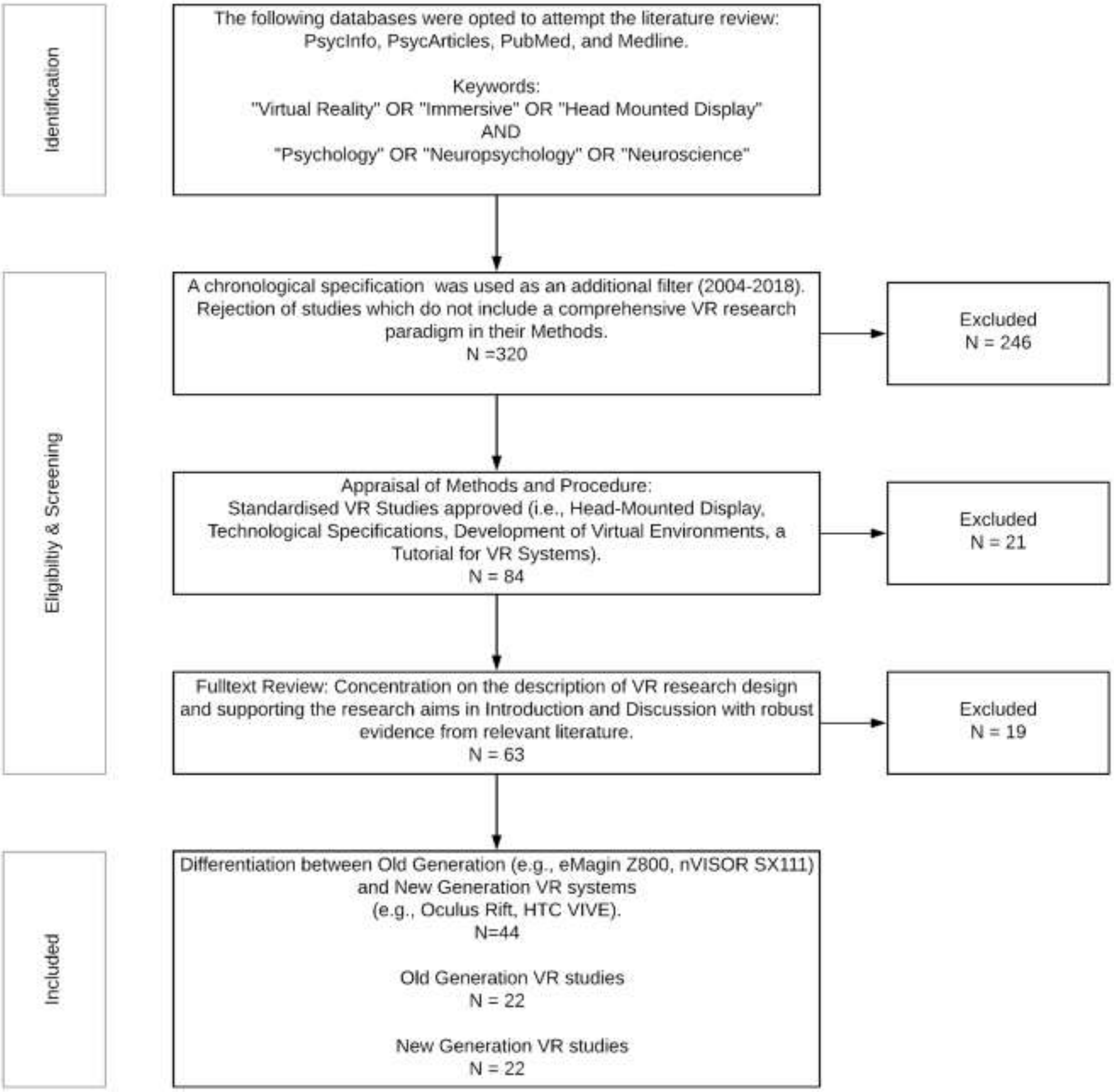

Table 4. Neuroscience studies employing old generation VR HMDs

| Study | Topic | HMD | Ergonomic Interactions | Clinical Condition | Age Group | N | VRISE | Dropouts |
|---|---|---|---|---|---|---|---|---|
| | | | YES =12<br>NO=10 | | | | YES=14<br>NO =8 | YES=9<br>NO=13 |
| Kim *et al.*, 2004 | Visuospatial Functions | Eye-trek FMD-250W | YES | Brain Injury | MA | 52 | YES | NO - 0 |
| Moreau *et al.*, 2006 | Executive functions | Eye-trek FMD-250W | YES | ADHD & Autism | YA | 22 | YES | YES - 1 |
| Botella *et al.*, 2007 | Therapy (VRET) | V6 VR | NO | Panic Disorder | YA & MA | 46 | YES | YES - 9 |
| Matheis *et al.*, 2007 | Memory | eMagin z800 | YES | Brain Injury | MA | 40 | NO | NO - 0 |
| Parsons *et al.*, 2007 | Executive functions | eMagin z800 | NO | ADHD | C | 20 | YES | YES - 1 |
| Banville *et al.*, 2010 | Memory | eMagin z800 | NO | Brain Injury | YA | 62 | NO | NO - 0 |
| Rizzo *et al.*, 2010 | Therapy (VRET) | eMagin z800 | NO | PTSD | YA | 20 | YES | YES - 5 |
| Reger *et al.*, 2011 | Therapy (VRET) | eMagin z800 | NO | PTSD | YA | 24 | YES | YES - 6 |
| Bioulac *et al.*, 2012 | Executive functions | eMagin z800 | YES | ADHD | YA | 36 | NO | NO - 0 |
| Carlozzi *et al.*, 2013 | Rehabilitation | eMagin z800 | YES | Spinal Cord Injury | MA | 54 | YES | YES - 10 |
| Meyerbroeker *et al.*, 2013 | Therapy (VRET) | nVISOR SX111 | YES | Agoraphobia | MA | 55 | YES | YES - 17 |
| Parsons *et al.*, 2013 | Attention Assessment | eMagin Z800 | YES | Healthy | YA | 50 | YES | NO - 0 |
| Peck *et al.*, 2013 | Racial Biases | nVISOR SX111 | YES | Healthy | YA | 60 | NO | NO - 0 |
| Freeman *et al.*, 2014 | Social Cognition | nVISOR SX111 | YES | Paranoia | YA | 60 | YES | NO - 0 |
| Rothbaum *et al.*, 2014 | Therapy (VRET) | eMagin Z800 | NO | PTSD | YA & MA | 156 | NO | NO - 0 |
| Veling *et al.*, 2014 | Paranoid Thoughts | eMagin Z800 | NO | Psychosis | YA | 41 | YES | NO - 0 |
| Hartanto *et al.*, 2014 | Social Stress | eMagin Z800 | NO | Healthy | YA | 54 | NO | NO - 0 |
| Gaggioli *et al.*, 2014 | Stress Levels | Vuzix Wrap 1200VR | NO | Healthy | MA | 121 | YES | NO - 0 |
| Shiban *et al.*, 2015 | Therapy (VRET) | eMagin Z800 | NO | Arachnophobia | YA | 58 | YES | YES - 8 |
| Freeman *et al.*, 2016 | Therapy (VRET) | nVISOR SX111 | YES | Persecutory Delusions | MA | 30 | YES | YES - 1 |
| Parsons & Carlew, 2016 | Attention Assessment | eMagin Z800 | YES | Healthy | YA | 50 | NO | NO - 0 |
| Parsons & Barnett, 2017 | Attention Assessment | eMagin Z800 | YES | Healthy | YA & OA | 89 | NO | NO - 0 |

*HMD = Head-Mounted Display; VRISE = VR induced adverse symptoms and effects; YA = Young Adults; MA = Middle-Aged Adults; OA = Older Adults; C = Children; VRET = VR Exposure Therapy; PTSD = Post-Traumatic Stress Disorder; ADHD = Attention Deficit Hyperactivity Disorder*

Table 5. Neuroscience studies employing new generation VR HMDs

| Study | Topic | HMD | Ergonomic Interactions | Clinical Condition | Age Group | N | VRISE | Dropouts |
|---|---|---|---|---|---|---|---|---|
| | | | YES =18<br>NO=4 | | | | YES=4<br>NO =18 | YES=4<br>NO=18 |
| Foerster *et al.*, 2016 | Attention Assessment | Oculus DK2 | NO | Healthy | YA | 44 | NO | YES - 2 |
| Quinlivan *et al.*, 2016 | Attention Assessment | Oculus DK2 | YES | Healthy | YA | 40 | NO | NO – 0 |
| Kim *et al.*, 2017 | VR Presence | Oculus DK2 | YES | PD | OA | 33 | NO | NO – 0 |
| Montenegro & Argyriou, 2017 | Memory, Attention, Executive Functions | Oculus DK2 | YES | AD (early stages) | OA | 20 | NO | NO – 0 |
| Parsons & McMahan, 2017 | Memory Assessment | HTC Vive | YES | Healthy | YA | 103 | NO | NO – 0 |
| Kelly *et al.*, 2017 | Spatial Perception | HTC Vive | YES | Healthy | YA | 76 | NO | NO – 0 |
| Bourdin *et al.*, 2017 | Fear of Death | Oculus DK2 | YES | Healthy | YA | 36 | NO | NO – 0 |
| Hasler *et al.*, 2017 | Racial Bias | Oculus DK2 | YES | Healthy | YA | 36 | NO | NO – 0 |
| Mottelson & Hornnaek, 2017 | Navigation, Attention, B-P | HTC Vive | YES | Healthy | YA & MA | 31 | NO | NO – 0 |
| Rooney *et al.*, 2017 | Social Cognition | Oculus Rift CV | YES | Healthy | YA & MA | 103 | NO | NO – 0 |
| Zimmer *et al.*, 2018 | Social Stress | Oculus DK2 | NO | Healthy | YA & MA | 93 | YES | YES - 5 |
| Hsieh *et al.*, 2018 | Spatial Perception & Navigation | HTC Vive | YES | Healthy | YA | 70 | NO | NO – 0 |
| Yeh *et al.*, 2018 | Anxiety | HTC Vive | YES | Healthy | YA | 34 | NO | NO – 0 |
| Collins *et al.*, 2018 | Psychoeducation on DBS | Oculus Rift CV | YES | Movement Disorder | OA | 30 | NO | NO – 0 |
| Barberia *et al.*, 2018 | Fear of Death | Oculus DK2 | YES | Healthy | YA | 31 | YES | YES - 1 |
| Banakou *et al.*, 2018 | Embodiment, Cognition – IQ | HTC Vive | YES | Healthy | YA | 30 | NO | NO – 0 |
| Christou *et al.*, 2018 | Motor-Rehabilitation | HTC Vive | YES | Stroke Patients | YA & MA | 29 | NO | NO – 0 |
| Gomez *et al.*, 2018 | Balance & Walking Rehabilitation | Oculus DK2 | YES | PD | OA | 22 | NO | NO – 0 |
| Lubetzky *et al.*, 2018 | Sensory Integration & Balance | Oculus DK2 | NO | Healthy | YA & MA | 21 | YES | NO – 0 |
| Oagaz *et al.*, 2018 | Memory Assessment | HTC Vive | YES | Healthy | YA | 20 | NO | NO – 0 |
| George *et al.*, 2018 | Working Memory & Attention Assessment | HTC Vive | YES | Healthy | YA | 20 | NO | NO – 0 |
| Detez *et al.*, 2019 | Gambling | HTC Vive | NO | Healthy | YA & MA | 60 | YES | YES - 3 |

*HMD = Head-Mounted Display; VRISE = VR induced adverse symptoms and effects; YA = Young Adults; MA = Middle-Aged Adults; OA = Older Adults; PD = Parkinson's disease; AD = Alzheimer's disease; DK = Development Kit; CV = Commercial Version; B-P = Body Perception; DBS = Deep Brain Stimulation*

## 3.2 Data Collection and Coding

### 3.2.1 Target Variables

The principal aim of the meta-analysis was to measure the frequency of VRISE in neuroscience or psychology studies using a VR HMD. However, only six studies reported VRISE quantitatively (i.e., using a questionnaire). For this reason, we considered only the presence or absence of VRISE. The dichotomous VRISE variable (i.e., presence or absence of VRISE) was quantified (i.e., absent VRISE = 0; present VRISE = 1) to facilitate a comparison (i.e., Bayesian t-tests) between the studies that used old generation HMDs, new generation DK HMDs, and new generation CV HMDs, as well as the examination of potential correlations with other variables (i.e., Bayesian Pearson's correlation analysis).

A secondary aim of the meta-analysis was to inspect the dropout rates in neuroscience or psychology studies that used VR HMDs. However, as the vast majority of studies had no dropouts, studies with some dropouts (e.g., 3, 5, 6) were statistically considered as outliers. For this reason, we considered the existence of dropouts in each study. The dropout variable was dichotomized as presence = 1 and absence = 0. This dichotomous dropout variable was used to investigate whether using a certain generation HMD (i.e., old generation HMDs, new generation DK HMDs, or new generation CV HMDs) could increase/decrease the dropout size. We compared (i.e., Bayesian t-tests) the dropout rate across studies that used old generation HMDs, new generation DK HMDs, and new generation CV HMDs. We also inspected whether the dropout rates correlated with other variables by using Bayesian Pearson's correlation analysis.

### 3.2.2 Grouping Variables

We subdivided studies into groups based on the HMD generation they used. Hence, two groups of studies were created and compared by using Bayesian t-tests; the first group included studies that utilized old generation HMDs, while the second group included studies which utilized new generation HMDs (i.e., both DKs and CVs).

The new generation studies were further distinguished and compared by using Bayesian t-tests based on the type of new generation HMDs adopted (i.e., DK or CV). Two sub-groups were formed; the first group included studies that utilized DK HMDs, and the second group included studies that utilized CV HMDs.

Furthermore, the recency of the HMD technology was compared by using an ordinal variable where 1 indicated old generation HMDs, 2 indicated new generation DKs, and 3 indicated new generation CVs. This ordinal variable allowed us to inspect whether the HMD generation correlated with other variables by using Bayesian Pearson's correlation analysis.

Lastly, we considered the type of interactions, where the type of interactions were expressed in a binary form (i.e., non-ergonomic interactions = 0 and ergonomic interactions = 1). This allowed a comparison between the VR studies which had ergonomic interactions and the VR studies which had non-ergonomic interactions by using a Bayesian t-test. It also allowed us to inspect whether the interaction type correlated with other variables (i.e., Bayesian Pearson's correlation analysis).

### 3.2.3 Definition of Ergonomic Interactions

In line with the definition of ergonomic interactions in our technological review, we considered interactions to be ergonomic or non-ergonomic based on their proximity to real-life interactions. We provide some examples below to clarify our criteria:

Example 1 - Ergonomic Interaction: if the VR software required the participant to look around moving his or her head.

Example 2 - Non-Ergonomic Interaction: if the VR software required the participant to look around by using a joystick or mouse.

Example 3 - Ergonomic Interaction: if the VR software required the user to interact with objects (e.g., pushing a button, holding an item) in the VE or to navigate within the VE by using either 6DoF controllers or direct-hand interactions.

Example 4 - Non-Ergonomic Interaction: if the VR software required the user to interact with objects (e.g., pushing a button, holding an item) in the VE or to navigate within the VE by using a keyboard or joystick (e.g., Xbox controller).

## 3.3 Statistical Analyses

Bayesian statistics were preferred over null hypothesis significance testing (NHST). The Bayesian factor ($BF_{10}$) was therefore used instead of p-values for statistical inference, although we do report both $BF_{10}$ and p-values. P-values measure the difference between the data and the null hypothesis ($H_0$) (e.g., the assumption of no difference or no effect), while the $BF_{10}$ calibrates p-values by converting them into evidence in favor of the alternative hypothesis ($H_1$) over the $H_0$ (Cox & Donnelly, 2011; Bland, 215; Held & Ott, 2018). $BF_{10}$ is considered substantially more parsimonious than the p-value in evaluating the evidence against the $H_0$ (Cox & Donnelly, 2011; Bland, 2015; Held & Ott, 2018). Also, the difference between $BF_{10}$ and the p-value in evaluating the evidence against $H_0$ is even greater in small sample sizes (Held & Ott, 2018). Bayesian Factor ($BF_{10}$) threshold $\geq 10$ was set for statistical inference in all analyses, which indicates strong evidence in favor of the $H_1$ (Rouder & Morey, 2012; Wetzels & Wagenmakers, 2012; Marsman & Wagenmakers, 2017), and corresponds to a p-value $<.01$ (e.g., $BF_{10} = 10$) (Cox & Donnelly, 2011; Bland, 2015; Held & Ott, 2018). JASP software was used to perform the statistical analyses (JASP Team, 2018). Bayesian independent samples t-tests were conducted to investigate the difference in VRISE frequency and dropout occurrence between old and new generation HMDs, as well as between new generation DKs and CVs. A Bayesian Pearson’s correlations analysis examined the possible statistical relationships amongst the HMD generations, VRISE presence, the type of interactions, and dropout occurrences.

## 3.4 Results

### 3.4.1 The Implementation of Old and New Generation HMDs in Cognitive Neuroscience

The studies that utilized old generation HMDs are displayed in Table 4 and recruited 1200 participants in total. Nine out of 22 studies examined stress disorders, 7 of these were VR exposure therapy (VRET) studies either for phobias or post-traumatic stress disorder (PTSD), while 2 studies attempted to assess stress levels in context (e.g., assessment of social stress during a job interview). In 9 studies, there were VR assessments of cognitive functions, 2 studies assessed memory, 3 studied attention, 3 examined executive functions, and one examined visuospatial ability. Two of the studies involved social cognition while only one involved paranoid thinking. Lastly, only one study provided rehabilitation sessions in VR for patients with spinal injuries. The targeted age groups were young adults in 18 studies, middle-aged adults in 8 studies, older adults in one study, and children in one study.

The studies that utilized new generation HMDs are displayed in Table 5 and recruited 982 individuals in total. Specifically, 376 individuals were recruited in 10 studies where new generation DKs were used, while 606 individuals were recruited in 12 studies where new generation CVs were used. Nine out of the 22 studies attempted to assess cognitive functions (i.e., memory, attention, visuospatial ability, executive functions), 4 investigated anxiety disorders (i.e., fear of death, social stress, general anxiety disorder), 3 provided sensorimotor rehabilitation interventions, 3 studies examined the effects of presence in specific VEs, 2 assessed social cognition and one study offered a psychoeducational session to patients with motor-related disorders. Lastly, the targeted age groups were young adults in 18 studies, middle-aged adults in 6 studies, and older adults in 4 studies.

### 3.4.2 Meta-Analysis

The descriptive statistics are presented in Figures 3 and 4. In Figure 3, the number of studies with VRISE are displayed according to their HMD generation and interaction type. In Figure 4, the dropouts and sample sizes are presented according to their HMD generation, VRISE presence, and interaction type. The presence of VRISE substantially becomes less frequent when new generation HMDs are implemented (Figure 3). In new generation HMDs, VRISE are present in only 4 out of 22 studies, while across 982 participants, there are only 11 dropouts. In contrast, in old generation HMDs, VRISE are present in 14 out of 22 studies, while in a total sample size of 1200 participants, there are 58 dropouts.

In the 14 old generation HMDs studies where VRISE are present, half of them involved ergonomic interactions and the other half involved non-ergonomic interactions. Similarly, there is an equal distribution of dropouts (29 in each) between the old generation HMDs studies that had ergonomic and non-ergonomic interactions. When only old HMDs with ergonomic interactions are considered, VRISE are present in 7 out of 12 studies, while in an entire sample size of 598, the dropouts are 29. In the studies with new generation DKs, non-ergonomic interactions had an increased presence of VRISE than the ones with ergonomic interactions. Also, in the studies which used DKs with non-ergonomic interactions, 7 participants out of 158 dropped out, while in studies with ergonomic interactions, only one participant out of 218 dropped out. Importantly, when new generation CVs with ergonomic interactions are exclusively considered, there are no VRISE or dropouts in any of the 11 studies with 546 participants. Finally, VRISE were only present in one study using a new generation CV HMD, where 3 participants dropped out. This study was the only one with a new generation CV HMD that did not involve ergonomic interactions.

The Bayesian independent samples t-test highlighted that studies involving new generation VR HMDs have significantly less frequent VRISE ($BF_{10} = 144.68$; $p < .001$). The difference in the existence of dropouts was not substantial, yet the studies with new generation HMDs have less frequent dropouts ($BF_{10} = 4.69$; $p < .05$) than studies with old HMDs. Notably, the studies which used a new generation CV HMD have significantly less frequent VRISE ($BF_{10} = 46.39$; $p < .001$) but not less frequent dropouts ($BF_{10} = 1.66$; $p = .16$) than the studies which used a new generation DK HMD. Finally, the studies which implemented VR software with ergonomic interactions had substantially less frequent VRISE ($BF_{10} = 19.54$; $p < .001$) and dropouts ($BF_{10} = 16.01$; $p < .001$) than studies which used VR software with non-ergonomic interactions.

The Bayesian Pearson's correlations demonstrated a substantial negative correlation between the presence of VRISE and the HMD generation ($BF_{10}$ = 328.03; r (44) = -0.56, p < .001), while the presence of VRISE robustly demonstrated a positive correlation with the existence of dropouts ($BF_{10}$ = 83510.53; r (44) = 0.68, p < .001). Also, the utilization of ergonomic interactions was significantly negatively correlated with VRISE ($BF_{10}$ = 20.11; r (44) = -0.42, p < .001) and the existence of dropouts ($BF_{10}$ = 16.11; r (44) = -0.41, p < .001).

**Figure 3. VRISE per HMD generation and ergonomic interactions.**

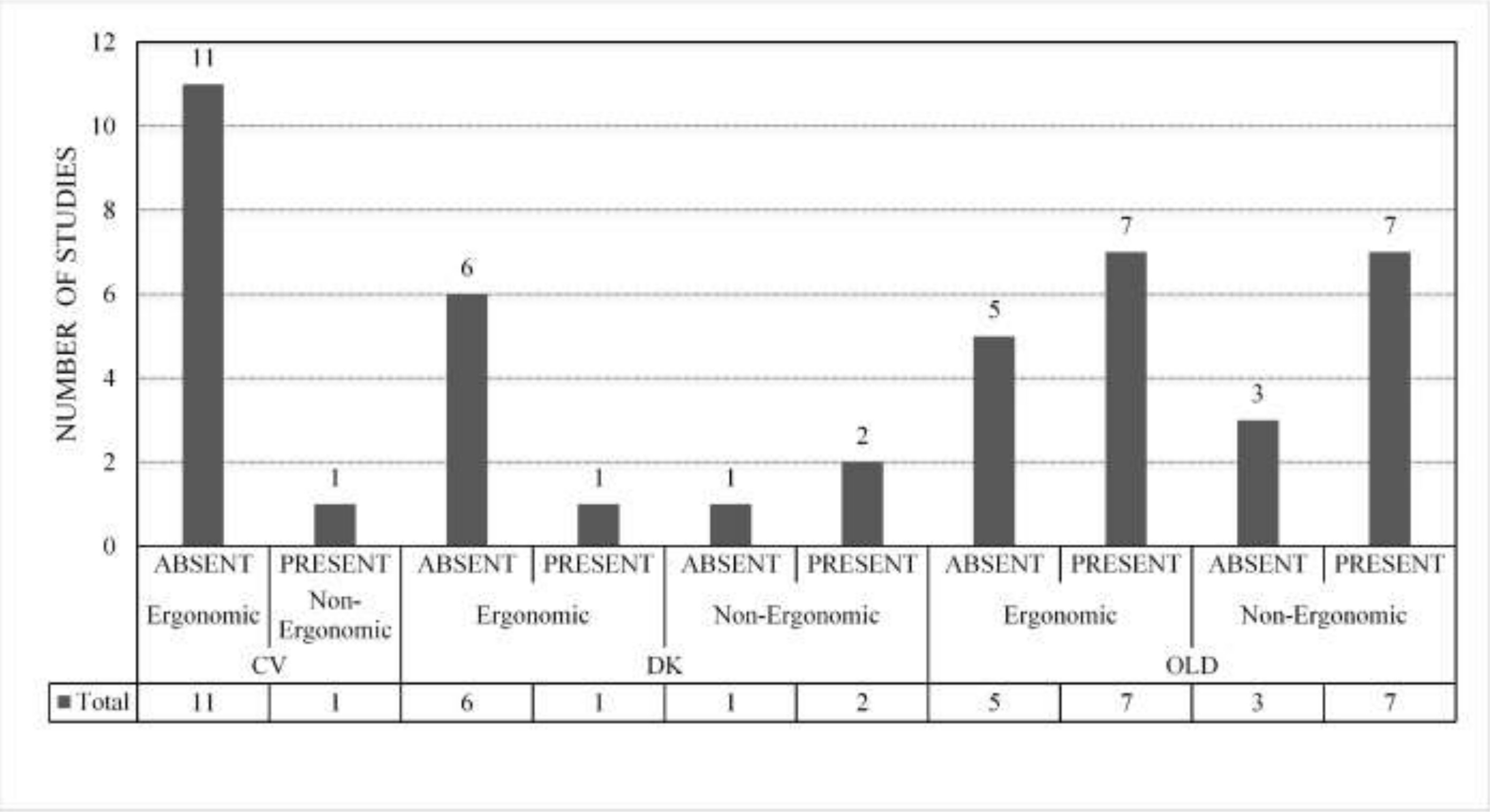

*ABSENT = Absence of VRISE; PRESENT = Presence of VRISE; OLD = Old Generation HMD; DK = New Generation HMD - Development Kit; CV = New Generation HMD – Commercial Version; Ergonomic = Ergonomic Interactions; Non-Ergonomic = Non-Ergonomic Interactions*

Figure 4. Dropouts and sample size per HMD, VRISE presence, and ergonomic interactions.

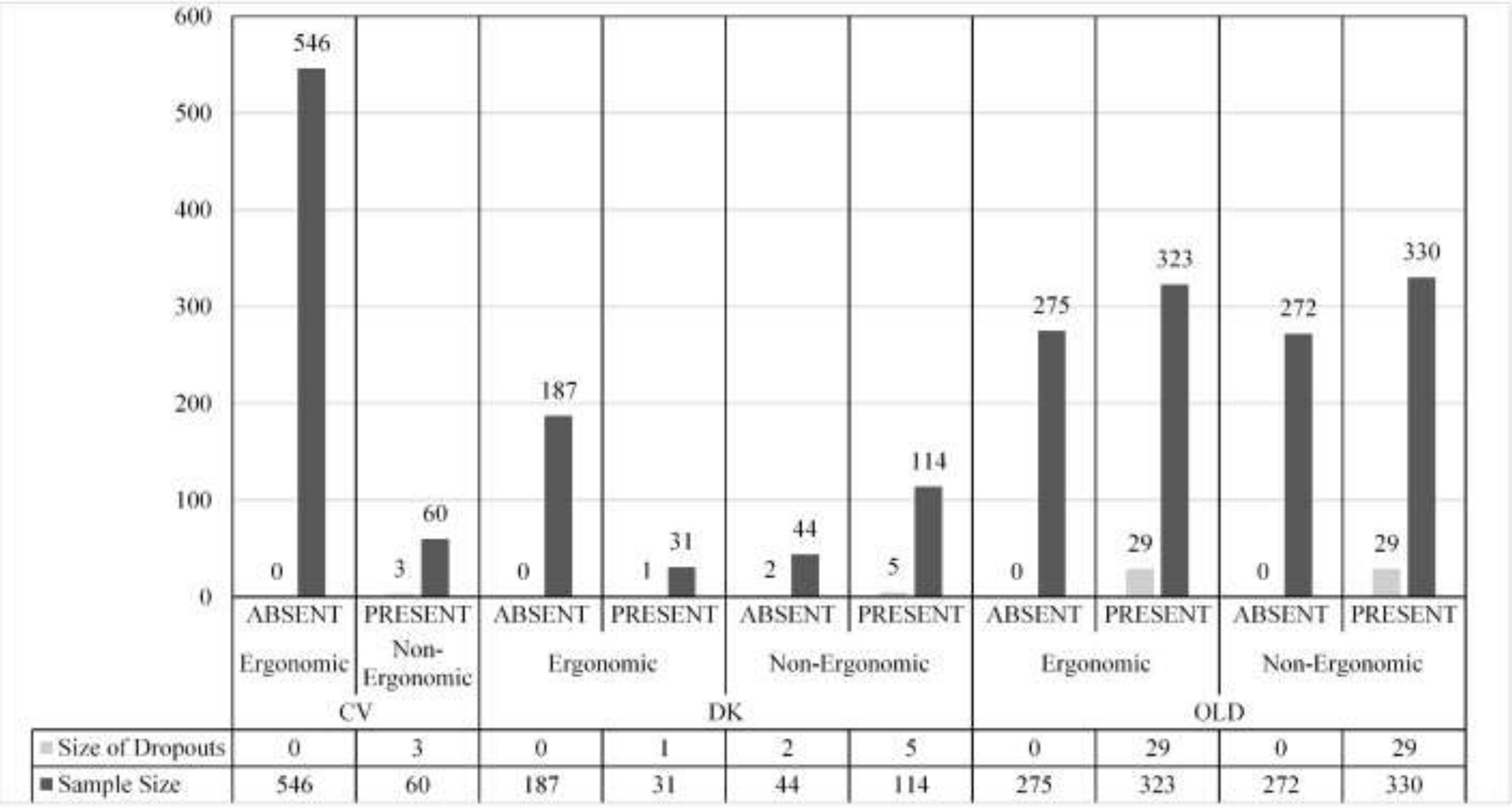

*ABSENT = Absence of VRISE; PRESENT = Presence of VRISE; OLD = Old Generation HMD; DK = New Generation HMD - Development Kit; CV = New Generation HMD – Commercial Version; Ergonomic = Ergonomic Interactions; Non-Ergonomic = Non-Ergonomic Interactions*

## 3.5 Discussion

The results of the meta-analysis indicated that VR HMDs have been implemented in diverse clinical conditions and age groups, as well as the unquestionable difference between old generation and new generation HMDs. There were significantly more frequent VRISE in the studies involving old generation VR HMDs compared to studies with new generation HMDs. Additionally, the frequency of VRISE correlated negatively with the HMD generation. Hence, the older the utilized HMD, the higher the VRISE frequency. Moreover, the existence of dropouts significantly and positively correlated with the presence of VRISE.

Nevertheless, one potential reason for the higher dropouts in old generation studies is that several studies included follow-up sessions (e.g., VRET) and participants may have opted not to return to complete the remaining sessions for reasons other than the presence of VRISE.  However, in the old generation studies, the dropout rates were low in relation to the size of the population, albeit there were VRISE present. The low dropout rates in the old generation HMD studies may be due to the fixed intervals between the VR sessions where the participants were able to rest and obtain relief from any adverse effects they were experiencing.

Furthermore, the incidence of VRISE in old generation HMD studies may be due to anxiety levels (Bouchard *et al.*, 2011) or be self-induced (Almeida *et al.*, 2017) as several of the studies had either stress-related aims or included participants with stress disorders. However, several of the new generation HMD studies also had comparable aims and/or populations and included patients with clinical conditions (e.g., Alzheimer's disease, Parkinson's disease, stroke, and movement disorders) which have high comorbidity with stress and anxiety (Factor *et al.*, 1995; Smith *et al.*, 2000; Jenner, 2003; Allen & Bayraktutan, 2009). Also, the rates of self-induced VRISE are expected to be equal in both new and old generation HMD studies. In addition, the reporting of VRISE may be for reasons unrelated to the quality of the hardware or software (e.g., subjectivity in the reporting of VRISE, individual differences in the experience of VRISE) (Kortum & Peres, 2014; Almeida *et al.*, 2017). However, this modulation is again expected to have affected both new and old generation HMD studies in a similar way.

Beyond the difference between old and new generations HMDs, a substantial difference is observed between DK and CV new generation HMDs. Significantly fewer VRISE were present in the studies that used a CV, indicating the superiority of new generation CV HMDs compared to new generation DK HMDs. Furthermore, the studies (i.e., both old and new generation studies) which utilized VR software with ergonomic interactions had robustly less frequent VRISE and dropouts than the studies which implemented VR software with non-ergonomic interactions. However, the ergonomic interactions do not appear to mitigate the dropout frequency and the incidence of VRISE in old generation HMDs. In contrast, VRISE were present in more DK studies with non-ergonomic interactions compared to DK studies with ergonomic interactions. Similarly, more participants dropped out from DK studies with non-ergonomic interactions. Notably, there were no VRISE or dropouts in CV studies with ergonomic interactions. Therefore, the contribution of ergonomic interactions in the reduction of VRISE increases when newer and better HMDs are utilized. To conclude, the findings of the meta-analysis are aligned with the outcomes of the technological review.

# 4 General Discussion

## 4.1 Technological Competence in VR Neuroscience and Neuropsychology

The findings of our technological literature review suggest that the hardware features of old generation HMDs and new generation DKs do not meet the minimum hardware features that alleviate or eradicate VRISE. Instead, the technological literature review postulates the suitability of new generation CVs which have specific hardware capabilities to alleviate VRISE. However, VR software attributes (e.g., ergonomic interactions) are equally vital.

Secondly, the findings of our meta-analysis of 44 neuroscientific or neuropsychological studies using VR are aligned with the outcomes of our technological review, where VRISE were substantially less frequent in studies which utilised new generation VR HMDs. In particular, the studies which used new generation CVs accompanied by ergonomic interactions did not have any VRISE or dropouts. Therefore, the combined outcomes of the technological review and the meta-analysis indicate that the appropriate VR HMDs are those with hardware characteristics equal to or greater than the HTC Vive and Oculus Rift, though the VR HMD should be implemented in conjunction with VR software which offers ergonomic interactions.

However, researchers may have to opt for an HMD based on their available budget. For example, the Oculus Rift costs around $400, while the HTC Vive costs around $500. Moreover, the majority of HMDs also require a VR-ready desktop PC or a laptop to be operated, so a researcher needs to additionally spend around $500-$1500 for a desktop computer or laptop to utilise these HMDs. Hence, the combined cost would be between $800 and $1900. The cost of VR equipment (e.g., both HMD and computer) may lead researchers to use HMDs that are cheaper, albeit that they are more likely to result in VRISE. However, in the market, there are plenty of cost-effective alternatives that meet the minimum hardware criteria. For example, the Oculus Quest is a standalone HMD (i.e., it does not require a PC, a laptop, or a smartphone to be operated) and it costs approximately $400. Hence, a researcher can spend the equivalent of the price of a neuropsychological test or a smartphone to acquire and use an HMD that meets the minimum hardware criteria to lower the presence of VRISE.

Nonetheless, the selection of an appropriate VR HMD and software requires technological competency from the researchers, clinicians, and/or research software developers. Unfortunately, the meta-analysis results do not indicate that technological knowledge of VR has been well established in neuroscience. Of course, the utilisation of old generation HMDs and new generation DKs pre-2016 is justified as the new generation CVs were not available. However, in our meta-analysis, 25 studies were conducted between 2016 and 2019, where half of these studies (13/25) implemented an inappropriate HMD (i.e., old generation HMD or new generation DK). However, 10 studies used a DK2 which has a marginally lower resolution than the minimum hardware criteria, while our meta-analysis results indicated that its utilisation in conjunction with ergonomic interactions appears to alleviate the frequency of VRISE, but not as effectively as the CV HMDs. Furthermore, one fifth of the studies did use a new generation HMD, but they did not have ergonomic interactions in their VR software. Therefore, at this time, VR technological competence does not seem to have been well established in neuroscience. As a result, in the studies since 2016, the health and safety of the participants may not substantially guaranteed, and the reliability of the results may be questionable, as VRISE substantially decreases reaction times and overall cognitive performance (Plant & Turner, 2009; Nalivaiko *et al.*, 2015; Plant, 2016; Nesbitt *et al.*, 2017; Mittelstaedt *et al.*, 2018), as well as confounding neuroimaging and physiological data (Toschi *et al.*, 2017; Arafat *et al.*, 2018; Gavgani et al, 2018). The selection of an appropriate HMD is paramount for successfully implementing VR HMDs in cognitive neuroscience and neuropsychology.

However, the implementation of the currently available and appropriate HMDs in neuroscience and neuropsychology should be compatible with the research aims. For example, in research designs where the user should be active (i.e., navigating, walking, and interacting within the VE) instead of being idle, or in a standing or a seated position, the researcher should opt for the best HMD that permits intense body movement and activity. In this setting, the Oculus Rift was found to be inferior to the HTC Vive on pick-and-place (i.e., relocating objects) tasks, whilst the HTC Vive also provided a substantially superior VR experience for users compared to the Oculus Rift (Suznjevic *et al.*, 2017). Moreover, the HTC Vive provides an interactive area that is twice the size (25 m2) of the Oculus Rift, albeit that both are very accurate in tracking (Borrego *et al.*, 2018). Nevertheless, the HTC Vive was found to lose motion-tracking and the ground level becomes slanted when the user goes out of bounds (Niehorster *et al.*, 2017). This shortcoming solely affects studies where the participant needs to go out of the tracking area. In most neuroscientific designs, the recommended maximum play area by HTC (6.25 $m^2$) or by Borrego *et al.*, 2018 (25 $m^2$) are both substantially adequate for conducting ecological valid experiments (Borrego *et al.*, 2018, Borges *et al.*, 2018). Nonetheless, the slanted floor or lost tracking is not a hardware problem but a software one (Borges *et al.*, 2018). In cases where the participant is required to go out of the tracking area, the tracking problem or the slanted floor may be easily corrected by adding 3 additional trackers (Peer *et al.*, 2018), using software with an improved algorithm (freely distributed by NASA Ames Research Centre) (Borges *et al.*, 2018), or by simply updating the firmware of the lighthouse base stations. In summary, the researchers should be technologically competent to not only identify and implement a safe HMD and software, but an HMD and software that facilitate the optimal research methods pertinent to their research needs and aims.

As discussed in our technological review, the quality of the implemented VR software is equally important to avoid VRISE. Our meta-analysis of VR studies indicated that the utilisation of ergonomic interactions is crucial albeit with the utilisation of an appropriate HMD. For example, Detez *et al.* (2019) used the HTC Vive to investigate physiological arousal and behaviour during gambling (Detez *et al.*, 2019). However, the interactions and navigation within the VE were facilitated by using a typical controller (Detez *et al.*, 2019). Hence, their VR software did not support the utilisation of the ergonomic 6DoF controllers (both hands) of the HTC Vive, which facilitate naturalistic navigation (e.g., teleportation) and interaction within the VE. Consequently, Detez *et al.*'s (2019) participants experienced VRISE (Detez *et al.*, 2019) and 3 of their participants discontinued their sessions and so their data were discarded (Detez *et al.*, 2019). Importantly, Detez *et al.* (2019) only reported the presence of VRISE and dropout size. They did not provide any quantitative data on the intensity of VRISE, or the quality of their software attributes (e.g., graphics, sound, tutorials, in-game instructions and prompts) (Detez *et al.*, 2019). Indeed, only six of the studies in the meta-analysis provided adequately explicit reports on VRISE and VR software. Since Detez *et al.* (2019) assessed reaction times and heart rates, these data are likely to be affected by VRISE, despite the study having a rigorous experimental design and using the HTC Vive. Therefore, it is important to use appropriate VR software and external hardware to prevent risks to the health and safety of the participants as well as the reliability of the results.

### 4.2 Limitations and Future Studies

The above technological review and meta-analysis of VR studies evidenced the importance of technological and methodological features in VR research and clinical designs. However, our meta-analysis of VR studies has some limitations. The meta-analysis considered VR studies with diverse populations and designs, which may have affected the frequency of VRISE and the existence of dropouts. Uniformity across studies (e.g., considering only VRET, assessments or a specific clinical population) was not possible due to the scarcity of neuroscience studies involving VR, especially using new generation HMDs. Moreover, the review did not consider any software details due to the scarceness of such descriptions in published studies. Future VR studies should report software and hardware features to allow an in-depth meta-analysis. Equally, only six studies provided quantitative reports of VRISE intensity, consequently, only the presence or absence of VRISE was considered. The dichotomous consideration of VRISE is susceptible to reports based on subjective criteria and individual differences, but this is likely to have affected the VRISE rates in both old and new generation studies. Future studies should aim to appraise the quality of the software and intensity of VRISE (e.g., using questionnaires). Studies should also attempt to clarify the acceptable duration of immersive VR sessions, which will aid researchers in designing their studies appropriately. Importantly, the cost of the VR software development should also be considered. Finally, studies should attempt to provide software development guidelines that enable researchers and/or research software developers to develop VR research software without depending on third parties (e.g., freelance developers or software development companies) and these guidelines should embed suggestions and instructions for VR software development, which meet the criteria discussed above.

### 4.3 Conclusion

The use of VR HMDs is becoming more popular in neuroscience either for clinical or research purposes and VR technology and methods have been well accepted by diverse populations in terms of age groups and clinical conditions. A more pleasant VR experience and a reduction in VRISE symptomatology has been found using new generation CV HMDs, which deliver an adequately high display resolution, rapid image refresh rate, ergonomic design and has controllers which allow naturalistic navigation and movement within the VE environment, especially when there is restricted teleportation. The outcomes of the current technological review and meta-analysis support the feasibility of new generation VR CV HMDs to be implemented in cognitive neuroscience and neuropsychology. The findings of the technological review suggest methods that should be considered in the development or selection of VR research software, as well as hardware and software features that should be included in the research protocol. The selected VR HMD and the VR research software should enable suitable ergonomic interactions, locomotion techniques (e.g., teleportation), and kinetic mechanics which ensure VRISE are reduced or completely avoided. A meticulous approach and technological competence are compulsory to consolidate the viability of VR research and clinical designs in cognitive neuroscience and neuropsychology.

## 5 Conflict of Interest

The authors declare that the research was conducted in the absence of any commercial or financial relationships that could be construed as a potential conflict of interest.

## 6 Author Contributions

The primary author had the initial idea and contributed to every aspect of this study. The rest of the authors contributed to the methodological aspects and the discussion of the results.

Table 1. Glossary of key terms and concepts

| | Terms & Concepts | Explanation/Definition |
|---|---|---|
| Headsets | Head Mounted Display (HMD) | A display device which is worn on the head and provides an immersive virtual reality for the wearer. |
| | Development Kit (DK) HMD | A prototype device, which is utilized by the VR Software developers to develop VR software before the commercial version of an HMD. The DKs are not provided for general use. |
| | Commercial Version (CV) HMD | The final version of an HMD, which is dispersed to the market for general use. |
| Display | Liquid Crystal Display (LCD) | A type of display/screen that uses the light-modulating properties of liquid crystals. Liquid crystals emit light indirectly, instead of using a reflector to produce images. |
| | Organic Light Emitting Diode (OLED) | A type of display/screen that uses an organic compound film that emits light in response to an electric current. OLEDs are used as displays in devices such as television screens, computer monitors, and smartphones. |
| | Field of View (FOV) | The area captured by the display device. The size of the FOV and the size of the display device directly affect the quality of the image. |
| | Refresh Rate & Frame Rate | The refresh rate is the number of times that the hardware updates its display per second. It involves the repeated display of identical frames. The frame rate indicates the frequency that software can add new data to a display. |
| | Resolution | The number of distinct pixels in each dimension displayed in a frame. |
| Interactions | Motion tracking | The process of tracking the movement of objects or people. It is facilitated by motion sensors which detect the position of motion trackers embedded in devices (e.g., HMDs and 6 DoF controllers). |
| | Controllers/Wands with 6 degrees of freedom (DoF) | Controllers which have 6DoF of movement in 3-D space on three directional axes (i.e., Forward-Back, Left-Right, Up-Down) and three rotational axes (i.e., Roll, Pitch, Yaw). |
| | Direct Hand Interaction | A motion tracking device (i.e., a motion sensor) which directly tracks hand movements |
| | Teleportation | A navigation system, which allows the user to be transferred to a new location in the virtual environment without physically moving in the real environment. |
| | Ergonomic Interactions | These resemble real-life interactions, which optimize user experience and overall VR system performance (see also 3.2.3 Definition of Ergonomic Interactions). Ergonomic interactions are facilitated by and restricted to the capabilities of the VR hardware and software. |
| | Virtual Environment (VE) | A three-dimensional artificial environment which is displayed on a display device and allows the users to interact with it. |

Table 2. Hardware specifications of old and new generation VR HMDs

| Product | Generation | Resolution (per eye) | Display Screen | Refresh Rate | FOV (Diagonal) | Motion Trackers and Sensors (Type and Quantity) |
|---|---|---|---|---|---|---|
| VFX 3D | Old | 480 x 240 | LCD | 45 Hz | 45° | - |
| VUZIX Wrap 1200 | Old | 852x480 | LCD | 60 Hz | 35° | Unknown type (1), 3 magnetometers, 3 accelerometers, and 3 gyroscopes |
| eMagin Z800 3DVisor | Old | 800x600 | OLED | 60 Hz | 40° | - |
| nVisor SX111 | Old | 1280x1024 | LCD | 60 Hz | 110° | - |
| Oculus Rift Development Kit 1 | New | 640 x 800 | LCD | 60 Hz | 110° | - |
| Oculus Rift Development Kit 2 | New | 960 x 1080 | OLED | 75 Hz | 110° | - |
| Oculus Rift Commercial Version | New | 1080 x 1200 | OLED | 90 Hz | 110° | Accelerometer, gyroscope, magnetometer, 360º constellation tracking camera. |
| HTC VIVE Commercial Version | New | 1080 x 1200 | OLED | 90 Hz | 110° | Sensors (>70) including MEMS, magnetometer, gyroscope, accelerometer, and laser position sensors, Lighthouse laser tracking system (2 base stations emitting pulsed InfraRed lasers), front-facing camera. |

*MEMS = Microelectromechanical systems*

Table 3. Criteria for a suitable VR Software in cognitive neuroscience and neuropsychology

| Domains | User Experience | Game Mechanics | In-Game Assistance | VRISE |
|---|---|---|---|---|
| CRITERIA | An Adequate Level of Immersion | A Suitable Navigation System (e.g., Teleportation) | Digestible Tutorials | Absence or Insignificant Presence of Nausea |
| | Pleasant VR Experience | Availability of Physical Movement | Helpful Tutorials | Absence or Insignificant Presence of Disorientation |
| | High Quality Graphics | Naturalistic Picking/Placing of Items | Adequate Duration of Tutorials | Absence or Insignificant Presence of Dizziness |
| | High Quality Sounds | Naturalistic Use of Items | Helpful In-game Instructions | Absence or Insignificant Presence of Fatigue |
| | Suitable Hardware (HMD and Computer) | Naturalistic 2-Handed Interaction | Helpful In-game Prompts | Absence or Insignificant Presence of Instability |

Table 4. Neuroscience studies employing old generation VR HMDs

| Study | Topic | HMD | Ergonomic Interactions | Clinical Condition | Age Group | N | VRISE | Dropouts |
|---|---|---|---|---|---|---|---|---|
| | | | YES =12<br>NO=10 | | | | YES=14<br>NO =8 | YES=9<br>NO=13 |
| Kim *et al.*, 2004 | Visuospatial Functions | Eye-trek FMD-250W | YES | Brain Injury | MA | 52 | YES | NO - 0 |
| Moreau *et al.*, 2006 | Executive functions | Eye-trek FMD-250W | YES | ADHD & Autism | YA | 22 | YES | YES - 1 |
| Botella *et al.*, 2007 | Therapy (VRET) | V6 VR | NO | Panic Disorder | YA & MA | 46 | YES | YES - 9 |
| Matheis *et al.*, 2007 | Memory | eMagin z800 | YES | Brain Injury | MA | 40 | NO | NO - 0 |
| Parsons *et al.*, 2007 | Executive functions | eMagin z800 | NO | ADHD | C | 20 | YES | YES - 1 |
| Banville *et al.*, 2010 | Memory | eMagin z800 | NO | Brain Injury | YA | 62 | NO | NO - 0 |
| Rizzo *et al.*, 2010 | Therapy (VRET) | eMagin z800 | NO | PTSD | YA | 20 | YES | YES - 5 |
| Reger *et al.*, 2011 | Therapy (VRET) | eMagin z800 | NO | PTSD | YA | 24 | YES | YES - 6 |
| Bioulac *et al.*, 2012 | Executive functions | eMagin z800 | YES | ADHD | YA | 36 | NO | NO - 0 |
| Carlozzi *et al.*, 2013 | Rehabilitation | eMagin z800 | YES | Spinal Cord Injury | MA | 54 | YES | YES - 10 |
| Meyerbroeker *et al.*, 2013 | Therapy (VRET) | nVISOR SX111 | YES | Agoraphobia | MA | 55 | YES | YES - 17 |
| Parsons *et al.*, 2013 | Attention Assessment | eMagin Z800 | YES | Healthy | YA | 50 | YES | NO - 0 |
| Peck *et al.*, 2013 | Racial Biases | nVISOR SX111 | YES | Healthy | YA | 60 | NO | NO - 0 |
| Freeman *et al.*, 2014 | Social Cognition | nVISOR SX111 | YES | Paranoia | YA | 60 | YES | NO - 0 |
| Rothbaum *et al.*, 2014 | Therapy (VRET) | eMagin Z800 | NO | PTSD | YA & MA | 156 | NO | NO - 0 |
| Veling *et al.*, 2014 | Paranoid Thoughts | eMagin Z800 | NO | Psychosis | YA | 41 | YES | NO - 0 |
| Hartanto *et al.*, 2014 | Social Stress | eMagin Z800 | NO | Healthy | YA | 54 | NO | NO - 0 |
| Gaggioli *et al.*, 2014 | Stress Levels | Vuzix Wrap 1200VR | NO | Healthy | MA | 121 | YES | NO - 0 |
| Shiban *et al.*, 2015 | Therapy (VRET) | eMagin Z800 | NO | Arachnophobia | YA | 58 | YES | YES - 8 |
| Freeman *et al.*, 2016 | Therapy (VRET) | nVISOR SX111 | YES | Persecutory Delusions | MA | 30 | YES | YES - 1 |
| Parsons & Carlew, 2016 | Attention Assessment | eMagin Z800 | YES | Healthy | YA | 50 | NO | NO - 0 |
| Parsons & Barnett, 2017 | Attention Assessment | eMagin Z800 | YES | Healthy | YA & OA | 89 | NO | NO - 0 |

*HMD = Head-Mounted Display; VRISE = VR induced adverse symptoms and effects; YA = Young Adults; MA = Middle-Aged Adults; OA = Older Adults; C = Children; VRET = VR Exposure Therapy; PTSD = Post-Traumatic Stress Disorder; ADHD = Attention Deficit Hyperactivity Disorder*

Table 5. Neuroscience studies employing new generation VR HMDs

| Study | Topic | HMD | Ergonomic Interactions | Clinical Condition | Age Group | N | VRISE | Dropouts |
|---|---|---|---|---|---|---|---|---|
| | | | YES =18 NO=4 | | | | YES=4 NO =18 | YES=4 NO=18 |
| Foerster *et al.*, 2016 | Attention Assessment | Oculus DK2 | NO | Healthy | YA | 44 | NO | YES - 2 |
| Quinlivan *et al.*, 2016 | Attention Assessment | Oculus DK2 | YES | Healthy | YA | 40 | NO | NO – 0 |
| Kim *et al.*, 2017 | VR Presence | Oculus DK2 | YES | PD | OA | 33 | NO | NO – 0 |
| Montenegro & Argyriou, 2017 | Memory, Attention, Executive Functions | Oculus DK2 | YES | AD (early stages) | OA | 20 | NO | NO – 0 |
| Parsons & McMahan, 2017 | Memory Assessment | HTC Vive | YES | Healthy | YA | 10 3 | NO | NO – 0 |
| Kelly *et al.*, 2017 | Spatial Perception | HTC Vive | YES | Healthy | YA | 76 | NO | NO – 0 |
| Bourdin *et al.*, 2017 | Fear of Death | Oculus DK2 | YES | Healthy | YA | 36 | NO | NO – 0 |
| Hasler *et al.*, 2017 | Racial Bias | Oculus DK2 | YES | Healthy | YA | 36 | NO | NO – 0 |
| Mottelson & Hornnaek, 2017 | Navigation, Attention, B-P | HTC Vive | YES | Healthy | YA & MA | 31 | NO | NO – 0 |
| Rooney *et al.*, 2017 | Social Cognition | Oculus Rift CV | YES | Healthy | YA & MA | 10 3 | NO | NO – 0 |
| Zimmer *et al.*, 2018 | Social Stress | Oculus DK2 | NO | Healthy | YA & MA | 93 | YES | YES - 5 |
| Hsieh *et al.*, 2018 | Spatial Perception & Navigation | HTC Vive | YES | Healthy | YA | 70 | NO | NO – 0 |
| Yeh *et al.*, 2018 | Anxiety | HTC Vive | YES | Healthy | YA | 34 | NO | NO – 0 |
| Collins *et al.*, 2018 | Psychoeducation on DBS | Oculus Rift CV | YES | Movement Disorder | OA | 30 | NO | NO – 0 |
| Barberia *et al.*, 2018 | Fear of Death | Oculus DK2 | YES | Healthy | YA | 31 | YES | YES - 1 |
| Banakou *et al.*, 2018 | Embodiment, Cognition – IQ | HTC Vive | YES | Healthy | YA | 30 | NO | NO – 0 |
| Christou *et al.*, 2018 | Motor-Rehabilitation | HTC Vive | YES | Stroke Patients | YA & MA | 29 | NO | NO – 0 |
| Gomez *et al.*, 2018 | Balance & Walking Rehabilitation | Oculus DK2 | YES | PD | OA | 22 | NO | NO – 0 |
| Lubetzky *et al.*, 2018 | Sensory Integration & Balance | Oculus DK2 | NO | Healthy | YA & MA | 21 | YES | NO – 0 |
| Oagaz *et al.*, 2018 | Memory Assessment | HTC Vive | YES | Healthy | YA | 20 | NO | NO – 0 |
| George *et al.*, 2018 | Working Memory & Attention Assessment | HTC Vive | YES | Healthy | YA | 20 | NO | NO – 0 |
| Detez *et al.*, 2019 | Gambling | HTC Vive | NO | Healthy | YA & MA | 60 | YES | YES - 3 |

*HMD = Head-Mounted Display; VRISE = VR induced adverse symptoms and effects; YA = Young Adults; MA = Middle-Aged Adults; OA = Older Adults; PD = Parkinson's disease; AD = Alzheimer's disease; DK = Development Kit; CV = Commercial Version; B-P = Body Perception; DBS = Deep Brain Stimulation*